\documentclass[english]{cimento}
\usepackage[T1]{fontenc}
\usepackage[latin9]{inputenc}
\usepackage{booktabs}
\usepackage{units}
\usepackage{amstext}
\usepackage{graphicx}

\makeatletter

\newcommand{\noun}[1]{\textsc{#1}}
\providecommand{\tabularnewline}{\\}

\instlist{\inst{ins:manc} Particle Physics Group, School of Physics and Astronomy, University of Manchester, UK}

\makeatother

\usepackage{babel}

\begin{document}
\title{Top Quark Spin Correlations at the Tevatron}

\author{Tim Head on behalf of the CDF and D0 collaborations%
\thanks{thead@fnal.gov%
}\from{ins:manc}}

\maketitle
\begin{abstract}
Recent measurements of the correlation between the spin of the top
and the spin of the anti-top quark produced in proton anti-proton
scattering at a centre of mass energy of $\sqrt{s}=\unit[1.96]{TeV}$
by the CDF and D0 collaborations are discussed. Using up to $\unit[4.3]{fb^{-1}}$
of data taken with the CDF and D0 detectors the spin correlation parameter
$C$, the degree to which the spins are correlated, is measured in
dileptonic and semileptonic final states. The measurements are found
to be in agreement with Standard Model predictions.
\end{abstract}

\section{Introduction}

Top quark physics at hadron colliders plays an important role in testing
the Standard Model of particle physics and its possible extensions.

In the Standard Model the top quark has a very short lifetime, $\tau_{\nicefrac{1}{2}}\approx\unit[5\times10^{\textrm{-}25}]{s}$,
therefore the definite spin state in which the top anti-top pair is
produced is not spoilt by hadronisation effects. As a result, the
direction of the spin of the top quark is reflected in the angular
distributions of its decay products. In contrast to this, the spin
of light quarks will flip before they decay, making the spin state
they are produced in unobservable. Furthermore, the theoretical calculations
necessary in order to predict the angular distributions can be performed
for top pairs, resulting in precise theoretical predictions which
can be tested by experiment. New physics in either the production
or decay mechanism would modify these angular distributions, making
spin correlations sensitive to new physics.

Until recently only one measurement of spin correlations has been
performed. Using $\unit[125]{pb^{\textrm{-}1}}$ of data taken during
Run~I of the Tevatron collider at Fermilab the D0 collaboration measured
a correlation coefficient in agreement with the Standard Model~\cite{Abbott:2000dt}.
However, since the sample contained only six events, the sensitivity
was too low to rule out the hypothesis of no spin correlations. Recently
the CDF and D0 collaborations performed measurements using up to $\unit[4.3]{fb^{\textrm{-}1}}$
of data taken with the CDF and D0~\cite{Abazov:2005pn} detectors,
the results of which are discussed below.

\section{Observables}

In strong interactions the top and anti-top quark are produced unpolarised
at hadron colliders, however the $t\bar{t}$ system is in a definite
spin state. At the Tevatron about $\unit[85]{\%}$, at next to leading
order, of top quark pairs are produced via quark anti-quark annihilation.
At threshold these $t\bar{t}$ systems will be in a $^{3}S_{1}$ state,
whereas the $\unit[15]{\%}$ of top quark pairs produced
via gluon fusion will be in a $^{1}S_{0}$ state. In the first case
the top and anti-top quark will tend to have their spins parallel,
in the second case they tend to be anti-parallel. One therefore expects
to observe a correlation between the direction of the spins.

The strength of the correlation due to the production mechanism can
be expressed as the asymmetry $A$,\begin{equation}
A=\frac{N_{\uparrow\uparrow}+N_{\downarrow\downarrow}-N_{\downarrow\uparrow}-N_{\uparrow\downarrow}}{N_{\uparrow\uparrow}+N_{\downarrow\downarrow}+N_{\downarrow\uparrow}+N_{\uparrow\downarrow}}\label{eq:production-asymmetry}\end{equation}
between the number of events with spins parallel, $N_{\uparrow\uparrow}$
and $N_{\downarrow\downarrow}$, and the number of events with spins
anti-parallel, $N_{\uparrow\downarrow}$ and $N_{\downarrow\uparrow}$.

\begin{figure}
\begin{centering}
\includegraphics[width=0.8\columnwidth]{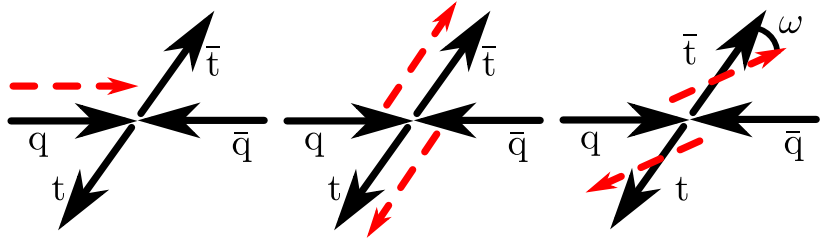}
\par\end{centering}

\caption{\label{fig:Three-spin-basis}The three choices of quantisation axis
used at the Tevatron. The {}``beamline basis'' (left) is optimal
for top pairs produced at threshold, the {}``helicity basis'' (centre) is
used for above threshold top pairs and the {}``helicity basis''
(right) interpolates between the two.}

\end{figure}

In order to measure the direction of the spin vector a quantisation
axis needs to be defined. At the Tevatron three sets of quantisation
axes, referred to as {}``spin basis'', are commonly used. They are
shown in Figure~\ref{fig:Three-spin-basis}.

The simplest is the so called {}``beamline basis'' in which the
direction of one of the incoming hadrons is used as quantisation axis.
This basis is easy to construct and is optimal for $t\bar{t}$ systems
produced at threshold. The production asymmetry has been calculated
at next to leading order (NLO) in QCD as $A=0.777$~\cite{Bernreuther:2004jv}.

The second basis is the {}``helicity basis'' in which the momentum
of the (anti)top quark in the top-anti-top quark zero momentum frame
is used to quantise the (anti)top quark spin. At the Tevatron the
strength of the correlation is smaller than in the {}``beamline basis'',
in NLO QCD $A=\textrm{-}0.352$. The opposite sign arises due to the
fact that the spins tend to be anti-parallel in this basis.

Finally the third basis is the {}``off-diagonal basis''. The direction
of the quantisation axes are defined by the angle $\omega$ with respect
to the (anti)top quark momentum. The angle $\omega$ is given by $\tan\omega=\sqrt{1-\beta^{2}}\tan\theta$,
where $\beta$ is the speed of the top quark and $\theta$ is its
scattering angle. This basis interpolates between the {}``beamline
basis'' close to threshold (low $\beta$) and the {}``helicity basis''
above threshold (large $\beta$). The production asymmetry is $A=0.782$.
While this is slightly larger than in the {}``beamline basis'' it
is more complex to reconstruct.

The angular distribution of decay product $i$ in the top quark rest
frame is given by:\begin{equation}
\frac{1}{\sigma}\frac{\textrm{d}\sigma}{\textrm{d}\cos\theta_{i}}=\frac{1}{2}\left(1-\alpha_{i}\cdot\cos\theta_{i}\right)\label{eq:top-decay}\end{equation}
where $\theta_{i}$ is the angle between the direction of flight of
decay product $i$ and the direction of the spin vector; $\alpha_{i}$
is the so-called spin analysing power. From Equation~\ref{eq:top-decay} it
is clear that the angular distribution of a decay product with $\alpha_{i}=0$
contains no information about the direction of the top quark spin
and the angular distribution of a decay product with $\alpha_{i}=\pm1$
will contain most information. The spin analysing power of the various
top quark decay products are listed in Table~\ref{tab:Spin-analysing-power}.
The particles with the highest spin analysing power are the lepton and the
down type quark from the W boson decay.

\begin{table}
\caption{\label{tab:Spin-analysing-power}Spin analysing power of the top quark
decay products. The up type quark, down type quark, neutrino and lepton
are the decay products of the W boson. For the antiparticles the sign
is reversed.}

\begin{tabular}{ccccc}
\toprule 
 & lepton, down type quark & neutrino & up type quark & b quark\tabularnewline
\midrule
\midrule 
analysing power $\alpha$ & +1 & +0.31 & +0.31 & +0.41\tabularnewline
\bottomrule
\end{tabular}
\end{table}

\begin{figure}
\begin{centering}
\includegraphics[width=0.55\columnwidth]{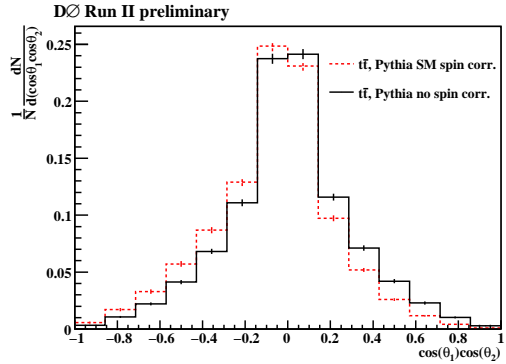}
\par\end{centering}

\caption{\label{fig:parton-coscos}The distribution $\cos\theta_{1}\cos\theta_{2}$
for a sample of top anti-top quark events using generated partons.
With spin correlations (red dashed) and without (black solid). Here
both top quarks decayed to leptons which subsequently were used as
spin analysers~\cite{d0dilep}.}

\end{figure}

In order to observe a correlation between the direction of the spin
of the top and anti-top quark one must consider the angle $\theta$
of a decay product of the top quark and the angle of a decay product
of the anti-top quark simultaneously. The double differential distribution
for a top quark decay product $i$ and anti-top quark decay product
$j$ is given by:\begin{equation}
\frac{1}{\sigma}\frac{\mathrm{d^{2}}\sigma}{\mathrm{d}\cos\theta_{i}\mathrm{d}\cos\theta_{j}}=\frac{1}{4}\left(1-A\alpha_{i}\alpha_{j}\cos\theta_{i}\cos\theta_{j}\right)\label{eq:coscos}\end{equation}
where $\sigma$ is the total cross section, $A$ is the production
asymmetry, and $\alpha_{i,\, j}$ is the spin analysing power of the
$i,\, j$-th decay product. In all analyses presented here the
spin correlation parameter $C=A\alpha_{i}\alpha_{j}$ is measured.
A measurement of the distribution given in Equation~\ref{eq:coscos}
should be performed as follows:
\begin{enumerate}
\item Reconstruct the top and anti-top quark momenta in the laboratory frame, 
\item Perform a boost from the laboratory frame to the rest frame of the
$t\bar{t}$ system. Define the vectors $\hat{b}_{i}$ and
$\hat{b}_{j}$ along which to quantise the top and anti-top quark
spins respectively.
\item Boost the top (anti-top) quark decay product to the top (anti-top)
quark rest frame and calculate $cos\theta_{i,\, j}=\hat{b}_{i,\, j}\cdot\hat{q}_{i,\, j}$.
\end{enumerate}
The difference between the case of no spin correlations, $A=0$, and
SM spin correlations as measured in the {}``beamline basis'', $A=0.777$,
using leptons as spin analysers is shown in Figure~\ref{fig:parton-coscos}.

\section{Measurements}

While in theory the down type quark is as powerful a spin analyser
as the lepton, it is more difficult to identify in practise. This
leads to two different approaches. In the first one, one selects a
pure sample of top pairs in which both the top and anti-top quark
decay to leptons. In the second a sample with higher statistics is
selected by requiring only one top quark to decay to a lepton. In
the following the advantages, challenges and results are discussed
for both approaches.

\subsection{Dilepton final states}

The advantages of the dilepton final state are that it is simple to
identify the final state particles of interest and the high purity
of the sample. The disadvantage is that one suffers from a low branching
ratio and needs to deal with two neutrinos when reconstructing the
kinematics of the event. Both CDF and D0 select events with two 
high $p_{t}$ leptons of opposite charge and at least two jets. The detailed
event selections are described in References~\cite{cdfdilep,d0dilep}
for CDF and D0, respectively. In final states with same flavour leptons
($e^{+}e^{\textrm{-}}$ and $\mu^{+}\mu^{\textrm{-}}$) the main background
arises from Drell-Yan, $Z\textrm{/}\gamma*\rightarrow\ell^{\textrm{-}}\ell^{\textrm{+}}$,
production. In the $e\mu$ final state the main background is instrumental,
this occurs mainly due to W+jets events in which a jet is misidentified
as a lepton. The second largest background arises due to semileptonic
decays of $Z\textrm{/}\gamma*\rightarrow\tau^{\textrm{-}}\tau^{\textrm{+}}$.
Further sources of background in all three final states
are the diboson processes WW, WZ and ZZ. Both signal and background
are modelled using Monte Carlo simulation, except for the instrumental
background which is estimated from data.

In order to reconstruct the momentum of the top and anti-top quark,
one needs to deal with the two neutrinos in the final state. To fully
characterise the kinematics of the final state one needs 18 quantities,
assuming the masses of the final state particles are known. While
the leptons and jets are observable in the detector, the two neutrinos
escape detection. It is possible to infer the sum of the momenta of
the neutrinos in the $x$ and $y$ plane from the missing transverse
energy, ${\displaystyle {\not}E_{T}^{x}}$ and ${\displaystyle {\not}E_{T}^{y}}$.
Using this information and making an assumption about the mass of
the W boson and the top quark it is possible to write down a set of
quartic equations which fully describe the final state. Solving them
yields up to four solutions per event. Additionally one needs to try
both lepton-jet pairings which increases the number of possible solutions
to eight.

\newpage
In the CDF measurement a likelihood function is constructed from several
observables and maximised with respect to the unknown neutrino momenta
($\vec{p}_{\nu}$, $\vec{p}_{\bar{\nu}}$) and the energies of the bottom
quark jets ($E_{b}^{\textrm{guess}}$, $E_{\bar{b}}^{\textrm{guess}}$):

\[
\begin{array}{l}
L\left(\vec{p}_{\nu},\,\vec{p}_{\bar{\nu}},\, E_{b}^{\textrm{guess}},\, E_{\bar{b}}^{\textrm{guess}}\right)=P\left(p_{z}^{t\bar{t}}\right)P\left(p_{T}^{t\bar{t}}\right)P\left(M_{t\bar{t}}\right)\times\\
\frac{1}{\sigma_{b}}\exp\left(-\frac{1}{2}\left(\frac{E_{b}^{\textrm{meas}}-E_{b}^{\textrm{guess}}}{\sigma_{b}}\right)^{2}\right)\times\frac{1}{\sigma_{\bar{b}}}\exp\left(-\frac{1}{2}\left(\frac{E_{\bar{b}}^{\textrm{meas}}-E_{\bar{b}}^{\textrm{guess}}}{\sigma_{\bar{b}}}\right)^{2}\right)\times\\
\frac{1}{\sigma_{x}^{\textrm{MET}}}\exp\left(-\frac{1}{2}\left(\frac{{\displaystyle \not}E_{x}^{\textrm{meas}}-{\displaystyle \not}E_{x}^{\textrm{guess}}}{\sigma_{x}^{\textrm{MET}}}\right)^{2}\right)\times\frac{1}{\sigma_{y}^{\textrm{MET}}}\exp\left(-\frac{1}{2}\left(\frac{{\displaystyle \not}E_{y}^{\textrm{meas}}-{\displaystyle \not}E_{y}^{\textrm{guess}}}{\sigma_{y}^{\textrm{MET}}}\right)^{2}\right)\end{array}\]
where $P\left(p_{z}^{t\bar{t}}\right)$, $P\left(p_{T}^{t\bar{t}}\right)$
and $P\left(M_{t\bar{t}}\right)$ are probability density functions
obtained from \noun{Pythia} $t\bar{t}$ Monte Carlo events, $E_{b,\,\bar{b}}^{\textrm{meas}}$
the measured energies of the bottom/anti-bottom quark jets, ${\displaystyle \not}E_{x,\, y}^{\textrm{meas}}$
the measured components of ${\displaystyle \not}E_{T}$, and $\sigma_{i}$
the respective resolutions. The maximisation is performed for both
lepton-jet pairings and the combination with the larger $L$ is kept.

As the \noun{Pythia }Monte Carlo simulation does not contain spin
correlations, templates for values of $C=-1,\,0.8,\dots,\,1$ are
obtained by reweighting the signal Monte Carlo at the generator level
using a weight $w\sim1-C\cdot\cos\theta_{1}\cos\theta_{2}$. For each
value of $C$ a two dimensional template in the decay angles of the
lepton and anti-lepton, $\cos\theta_{\ell^{+}}$ , $\cos\theta_{\ell^{\textrm{-}}}$
and a template in the decay angles of the bottom quark jets, $\cos\theta_{b}$
, $\cos\theta_{\bar{b}}$ is created. The two templates are fit with
an analytical function $f^{\ell,\, b}\left(x,\, y;\, C\right)$. The
measurement is then performed on the $N$ candidate events by maximising
the likelihood function: \[
L\left(C\right)=\prod_{i=0}^{N}f^{l}\left(x,\, y;\, C\right)f^{b}\left(x,\, y;\, C\right).\]

\begin{figure}
\begin{centering}
\includegraphics[width=0.8\columnwidth]{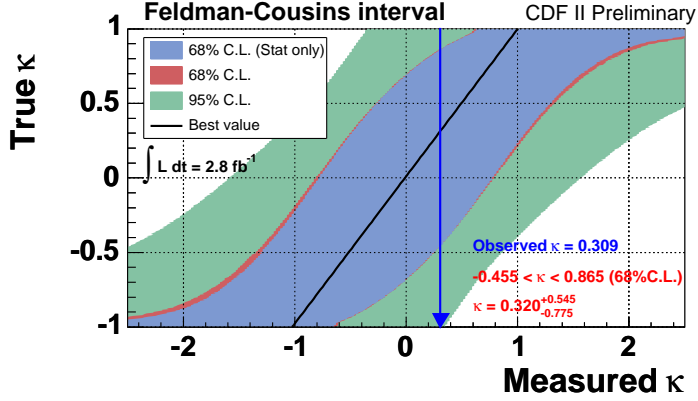}
\par\end{centering}

\caption{\label{fig:FC-belts}The 68\% (stat only), 68\% and 95\% Confidence
Level intervals constructed according to the Feldman-Cousins prescription
including statistical and all systematic uncertainties for the CDF
measurement. The best fit value is $C=0.32_{-0.78}^{+0.55}$~\cite{cdfdilep}.}

\end{figure}

In order to extract limits from the measurement, a confidence belt
according to the Feldman-Cousins prescription \cite{Feldman:1997qc}
is created. This naturally includes both statistical and systematic
uncertainties and allows one to decide before looking at the data
whether to quote a one or two sided limit. Using $\unit[2.8]{fb^{\textrm{-}1}}$
of data the best fit value is $C=0.32_{-0.78}^{+0.55}\textrm{(stat + syst)}$
and the corresponding confidence belts are shown in Figure~\ref{fig:FC-belts}.
The measurement was performed in the {}``helicity basis''. The result
is consistent with the expected value of $C=0.782$. The largest contributions
to the systematic uncertainty come from evaluating the PDF uncertainties
and the finite number of Monte Carlo events used to form the templates.

At D0 the neutrino weighting technique is used to solve for the event
kinematics. By making an assumption about the rapidity, $\eta$, of
the neutrino and anti-neutrino, it is possible to solve the event
kinematics, while not using ${\displaystyle {\not}E_{T}^{x}}$ and
${\displaystyle {\not}E_{T}^{y}}$ in the process but instead to assign
a weight, $w$, to each solution given by:\[
w=\exp\left(-\frac{\left({\displaystyle {\not}E_{T}^{x}-\nu_{x}-\bar{\nu}_{x}}\right)^{2}}{\sigma^{2}}\right)\times\exp\left(-\frac{\left({\displaystyle {\not}E_{T}^{y}-\nu_{y}-\bar{\nu}_{y}}\right)^{2}}{\sigma^{2}}\right)\]
where $\nu_{x,\, y}$ and $\bar{\nu}_{x,\, y}$ are the x and y components
of the neutrino and anti-neutrino momentum for a given solution and
$\sigma$ is the ${\not}E_{T}^{x}$ resolution. Many solutions are
obtained by sampling the neutrino and anti-neutrino rapidity based on
Monte Carlo simulation. No dependence of the neutrino rapidity on the
presence of spin correlations is observed. The weighted mean of all
solutions for an event is used as estimator for the true value of
$\cos\theta_{\ell^{+}}\cos\theta_{\ell^{-}}$.

As for the CDF measurement, the \noun{Pythia} Monte Carlo simulation
is used to model the signal sample. A one dimensional template in
the variable $\cos\theta_{\ell^{+}}\cos\theta_{\ell^{-}}$ is created
for $C=0$ and $C=0.777$ by reweighting the distribution at the generator
level. In order to extract a value of $C$ a linear combination of
the two templates is fit to the data.

Pseudo-experiments are created for each value of $C$ and fit with
signal and background templates. Each source of systematic uncertainty
is considered as a nuisance parameter during the fit. Feldman-Cousins
confidence belts are constructed from the pseudo experiments. Using
up to $\unit[4.2]{fb^{\textrm{-}1}}$ of data the best fit value is
$C=-0.17_{-0.53}^{+0.64}\textrm{(stat + syst)}$. In this measurement
the {}``beamline basis'' was used and the measured value is consistent
with the Standard Model expectation of $C=0.777$ at the two sigma
confidence level.

The two main sources of systematic uncertainty are the variation of
the assumed top mass during the event reconstruction from $\unit[175]{GeV}$
to $\unit[170]{GeV}$ and the test of the reweighting method. For
the latter, the two \noun{Pythia} signal templates were replaced by
\noun{Alpgen}, which contains spin correlations, and MC@NLO where
spin correlations were turned off.

\begin{figure}
\begin{centering}
\includegraphics[width=0.4\columnwidth]{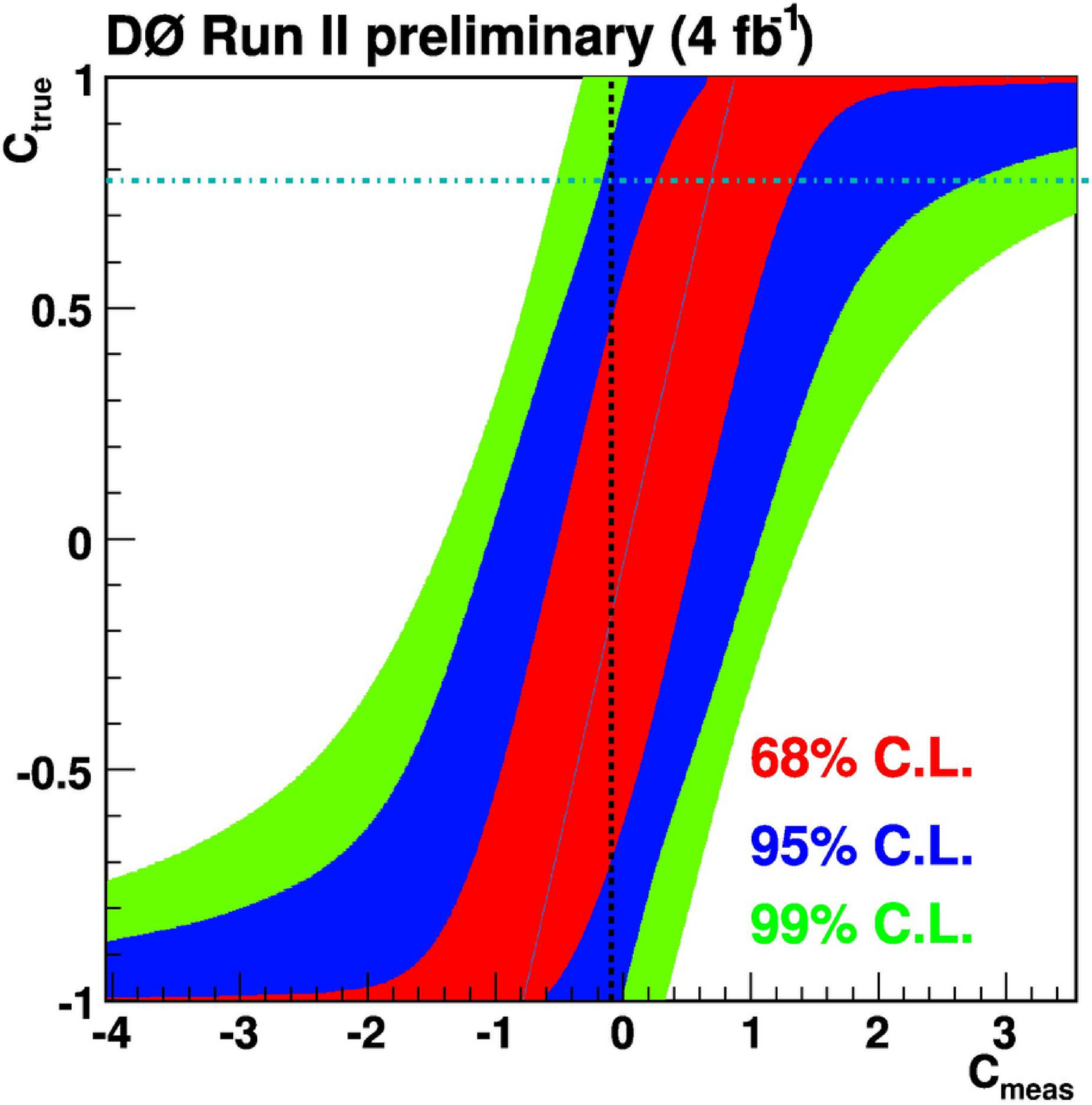}\includegraphics[width=0.5\columnwidth]{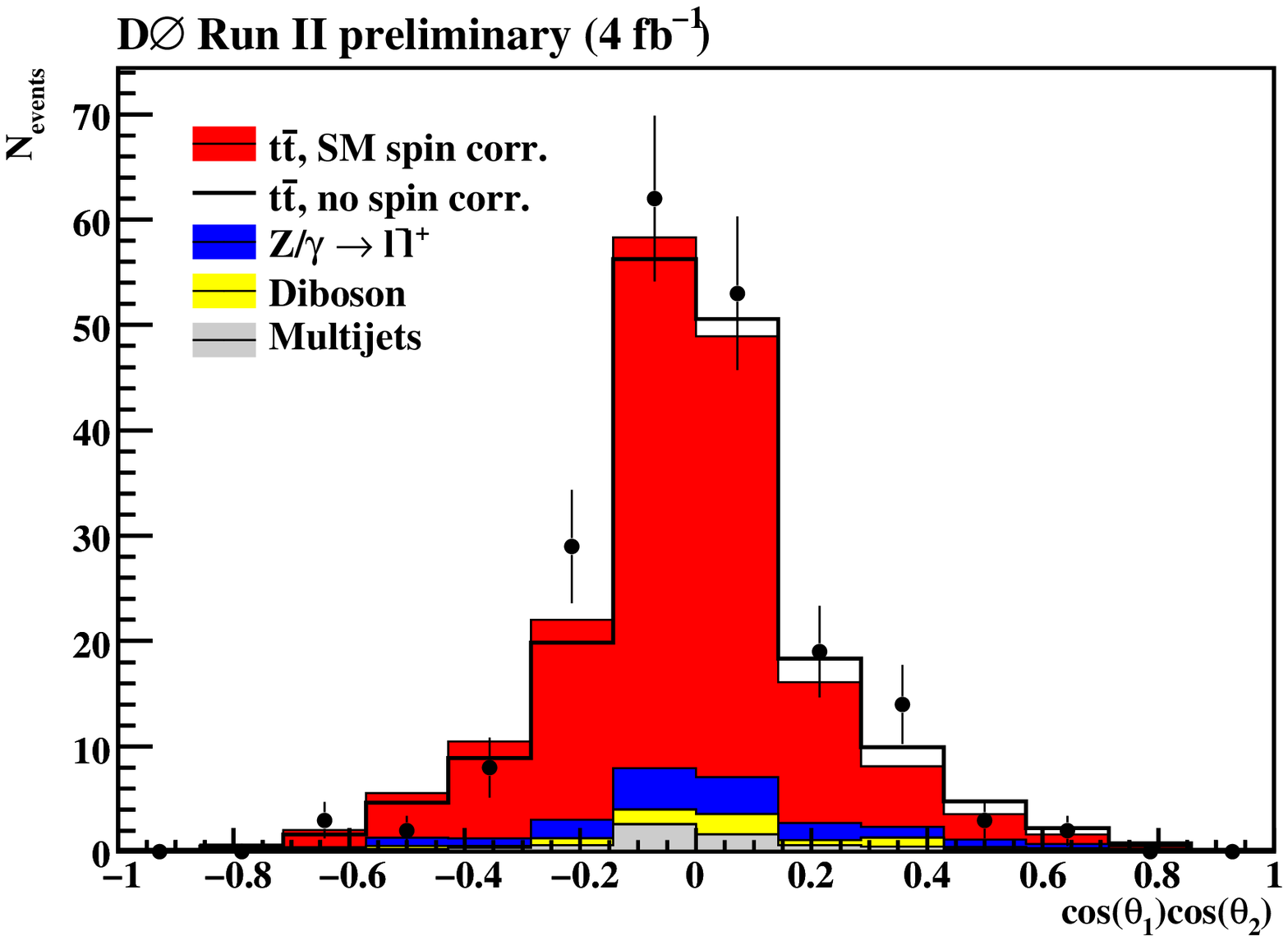}
\par\end{centering}

\caption{Left: the 68\%, 95\% and 99\% Feldman-Cousins confidence belts are
shown. The best fit value can be read of at the intersection of the
dashed black line and the thin blue line. Right: The sum of all dilepton
channels is shown. The open black histogram shows the expected distribution
for the case of no spin correlations, $C=0$ and the filled red histogram
the expected distribution for Standard Model spin correlations, $C=0.777$~\cite{d0dilep}.}

\end{figure}

\subsection{Semileptonic final states}

Selecting semileptonic events results in a higher yield, but the challenge
is to identify the down type quark. This is done probabilistically
by choosing the jet closest to the bottom type jet in the W boson
rest frame \cite{Mahlon:1995zn}, which will result in picking the
correct jet about 60\% of the time.

Events are selected by requiring at least one high $p_{T}$, central
lepton, large missing transverse energy and four
or more jets, one of which must be identified as a b-jet. The backgrounds
are estimated both from simulation and data. For details of the selection
see Reference~\cite{cdfljets}. Using $\unit[4.3]{fb^{\textrm{-}1}}$
of data a total of 1001 events are selected of which 786 are expected
to be top pair events.

When produced in pairs the top and anti-top quark either have the
same helicity or opposite helicity. The fraction of top pairs with
opposite helicity is given by:\[
f_{O}=\frac{\sigma\left(\bar{t}_{R}t_{L}\right)+\sigma\left(\bar{t}_{L}t_{R}\right)}{\sigma\left(\bar{t}_{R}t_{R}+\bar{t}_{L}t_{L}+\bar{t}_{R}t_{L}+\bar{t}_{L}t_{R}\right)},\]
where $\sigma\left(\bar{t}_{L,\, R}t_{L,\, R}\right)$ denotes the
cross section for each possible helicity configuration. Using Equation~\ref{eq:production-asymmetry}
one can show that a measurement of $f_{O}$ is equivalent to a measurement
of $A$ in the helicity basis.

One template for top pairs with same helicity and one template for
top pairs of opposite helicity are created using a modified version
of the \noun{Herwig} event generator. The opposite helicity fraction
is extracted with a binned maximum likelihood fit of the two templates
to the data, with contributions from backgrounds taken into account.
The best fit value is $f_{O}=0.80\pm0.26\textrm{(stat + syst)}$ or
equivalently $A=2f_{O}-1=0.60\pm0.52\textrm{(stat + syst)}$. This
is consistent with the Standard Model expectation of $A=0.4$. The
two main systematic uncertainties are Monte Carlo statistics and jet
energy scale.

\begin{figure}
\begin{centering}
\includegraphics[width=0.45\columnwidth]{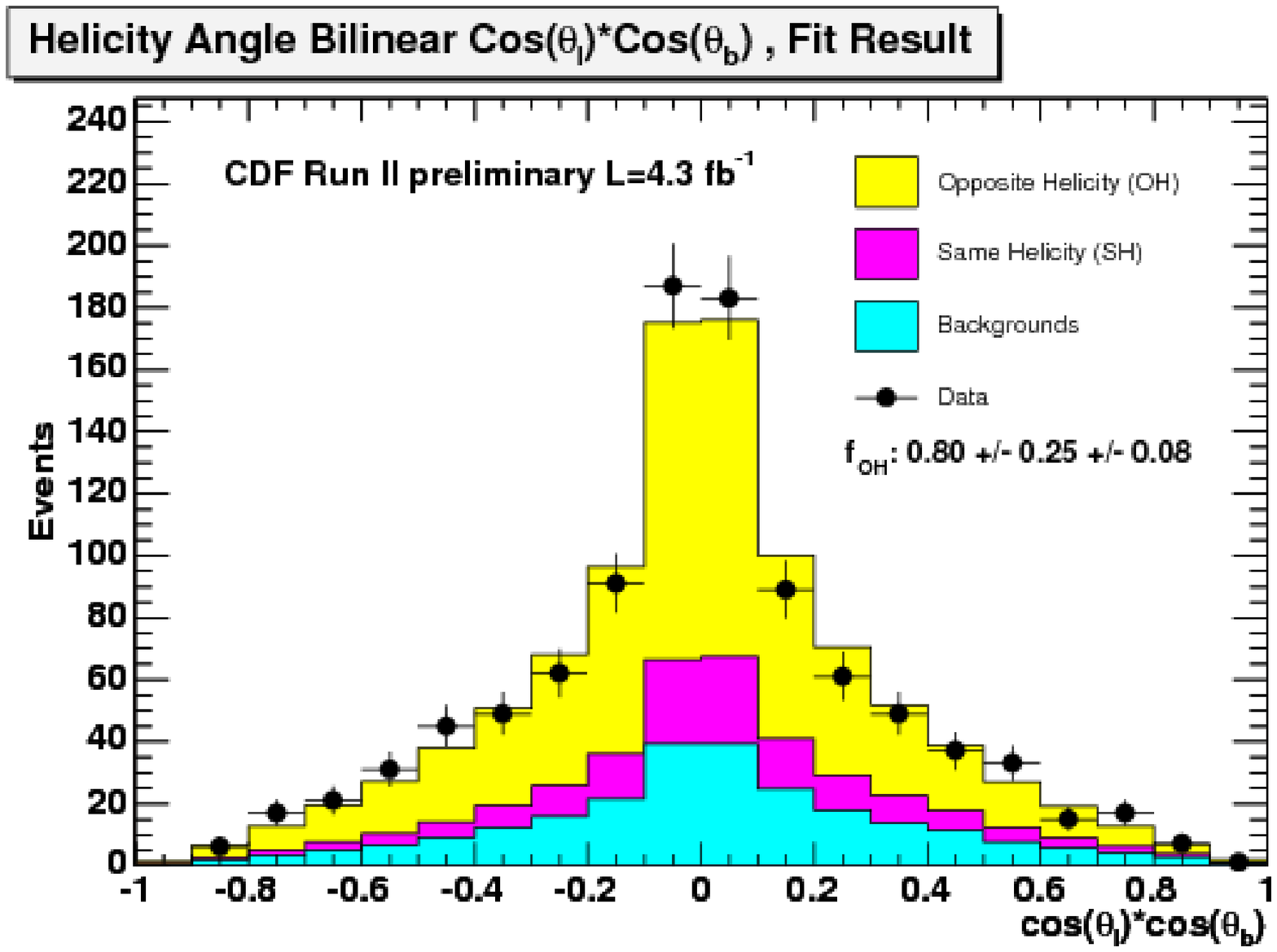}\includegraphics[width=0.45\columnwidth]{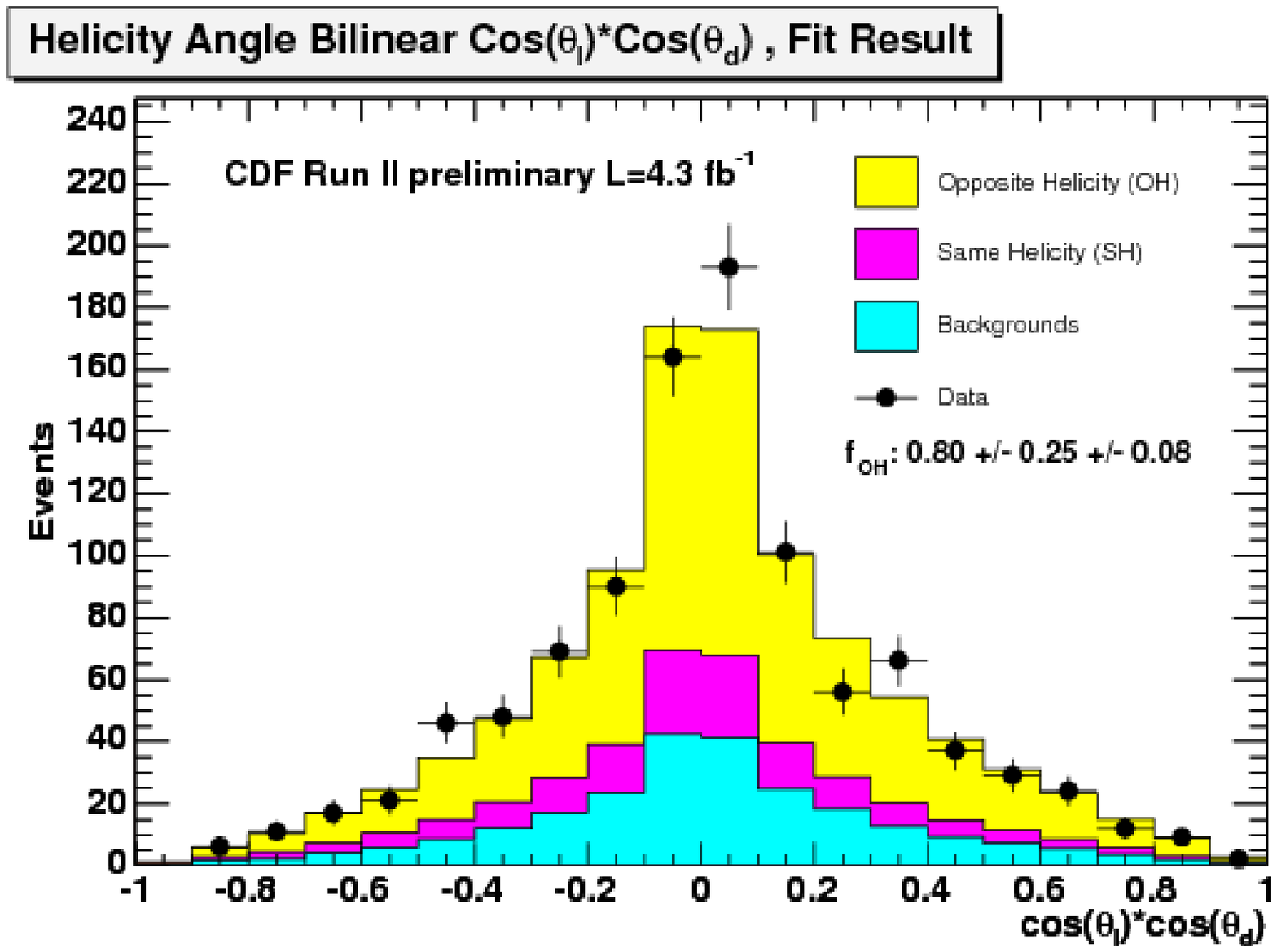}
\par\end{centering}

\caption{The best fit of same helicity, opposite helicity and background templates
for the CDF semileptonic decay channel. On the (left) the distribution
of the product of the decay angle of the lepton and the bottom quark.
On the (right) the distribution of the product of the decay angle
of the lepton and the down type quark. The best fit value from a simultaneous
fit to both distributions is $f_{O}=0.8\pm0.26\textrm{(stat + syst)}$
or $C=0.6\pm0.52\textrm{(stat + syst)}$~\cite{cdfljets}.}

\end{figure}

\section{Conclusions}

The spin correlation parameter $C$ has been measured in dilepton
and semileptonic decays of top and anti-top quark pairs using up to
$\unit[4.3]{fb^{-1}}$ of data collected with the CDF and D0 detectors.
Measurements were performed in the {}``beamline'', {}``helicity''
and {}``off-diagonal'' bases. The measurements are found to be in
agreement with the Standard Model predictions. All three measurements
are still statistically limited. Considering that the Tevatron collider
has delivered nearly twice as much integrated luminosity since the
analyses have been performed, updates of all measurements can be expected
soon.

\bibliographystyle{varenna}
\bibliography{thead}

\end{document}